
\rightline{IASSNS 93/64}
\bigskip
\centerline{\bf Measurement Theory in Lax-Phillips Formalism}
\bigskip
\bigskip

\centerline{
S. Tasaki ${\dag}$
\footnote{${}^{\ast}$}{Present address: Institute for Fundamental
Chemistry, 34-4 Takano-Nishihiraki-cho, Sakyo-ku, Kyoto 606, Japan.},
E. Eisenberg $\S$ and
L.P. Horwitz $\S \ddag$ \footnote{${}^{\natural}$}{Permanent
address: School of Physics,  Tel Aviv University, Ramat Aviv, Israel}
}

\bigskip
\centerline{${\dag}$ International Solvay Institutes for Physics and Chemistry}
\centerline{CP 231 Campus Plaine ULB, Boulevard de Triomphe}
\centerline{1050 Brussels, Belgium}
\centerline{$\S$ Department of Physics, Bar-Ilan University}
\centerline{Ramat-Gan 52900, Israel}
\centerline{$\ddag$ Institute for Advanced Study, School of Natural Sciences}
\centerline{Princeton, NJ 08540}

\vfil

\bigskip\noindent
{\it Abstract.\/}
It is shown that the application of Lax-Phillips scattering theory to
quantum mechanics provides a natural framework for the realization of the
ideas of the Many-Hilbert-Space theory of Machida and Namiki to describe the
development of decoherence in the process of measurement.  We show that if
the quantum mechanical evolution is pointwise in time, then decoherence
occurs only if the Hamiltonian is time-dependent.  If the evolution is
not pointwise in time (as in Liouville space), then the decoherence
may occur even for closed systems.  These conclusions apply as well to the
general problem of mixing of states.

\eject
\bigskip\noindent
{\bf 1. Introduction}
\bigskip
Many attempts have been made for explaining quantum measurements [1].
Recently, Machida and Namiki have proposed a measurement theory called
Many-Hilbert-Space (MHS) theory [2,3,4].  They observed that the
explanation of quantum mechanical measurements only requires
decoherence of the wave function, which can be formulated dynamically
by extending the framework of quantum mechanics.  In their theory,
the key role is played by a direct integral space of continuously
many Hilbert spaces and a continuous superselection rule.  The direct
integral space structure is assigned only to the measurement apparatus
reflecting its macroscopic nature, while the observed system remains
to be described by a single Hilbert space. Various situations such as
double slit experiments and negative result experiments have been
investigated in the framework of the MHS theory [3,4].

A direct integral space of continuously many Hilbert spaces appears
naturally in the quantum version of the Lax-Phillips theory [5]
introduced by Flesia, Piron and Horwitz [6,7].  In this approach, the
direct integral space is introduced in order to allow the generator of
motion to have a spectrum over the whole real axis, which is a
necessary condition of the application of the Lax-Phillips theory.
Contrary to the MHS theory, the direct integral space structure is
assigned to every system irrespective to its size.  Therefore, the
Flesia-Piron-Horwitz approach may provide a measurement theory which
inherits the desired features of the MHS theory and which does not
need the clear-cut boundary between the system and the apparatus.  In
this paper, we show that, in fact, the quantum Lax-Phillips theory as
constructed by Flesia and Piron admits the possibility of dynamical
decoherence of states, and therefore may be used for the description
of the measurement process.  Futhermore, we show that using an
extension of the Flesia-Piron form of the theory, which was shown [8]
to be necessary for achieving the full structure of Lax-Phillips
theory (i.e., non-trivial S-matrix), decoherence may occur even for
{\it closed} (homogenous) systems.

We first introduce the Lax-Phillips  description of quantum mechanics starting
from the Schr\"odinger representations.  Consider a quantum mechanical
system with a Hilbert space $\cal H$.  In the Lax-Phillips  formalism
introduced
in refs. [6,7] the state space is described by a direct integral of
(isomorphic) copies ${\cal H}_t$ of the Hilbert space $\cal H$, indexed
by $t$

$$ \overline{\cal H} \equiv \int_\oplus {\cal H}_t dt.
\eqno(1)$$
Scalar products in $\bar{\cal H}$ have the form

$$(f,g) = \int (f_t, g_t)_{\cal H} dt,
\eqno(2)$$
and the norm squared in

$$\parallel f \parallel^2_{\bar {\cal H}} = \int \parallel f_t
\parallel^2_{\bar {\cal H}} dt.
\eqno(3)$$

We assume the existence of a unitary group of evolution operators
parameterized by the laboratory time $\tau$ (which is not a dynamical
variable of the system, but only a parameter) $U(\tau)$ in ${\cal H}$, with an
infinitesimal generator $K$.

In the Flesia-Piron-Horwitz approach [6,7], the evolution $U(\tau)$ is
expressed by an operator $W_t (\tau)$ local in the index $t$:

$$ (U(\tau) \psi)_{t+\tau} \equiv \psi^\tau_{t+\tau} = W_t(\tau)
\psi_t
\eqno(4)$$
with the generator

$$ (K \psi)_t = -i \partial_t \psi_t + H(t) \psi_t,
\eqno(5)$$
where $H(t)$ corresponds to the Hamiltonian in the usual Hilbert
space ${\cal H}$.

It was recently shown [8] that in order to achieve the full structure
of the Lax-Phillips  theory (i.e., a non-trivial S-matrix which can be related
to the Lax-Phillips  semigroup), one must consider a more general evolution law
of the form

$$ (U(\tau) \psi)_{t+\tau} = \int^{+\infty}_{-\infty} W_{t,t^\prime} (\tau)
\psi_{t^\prime} dt^\prime;
\eqno(6)$$
the generator of the evolution in this case takes the form

$$ (K \psi)_t = -i \partial_t \psi_t + \int^{+\infty}_{-\infty}
\kappa_{t,t^\prime} \psi_{t^\prime} dt^\prime,
\eqno(7)$$
We remark that the structure of the evolution in the Liouville space
[10] has precisely this form [8], if we take the $t$-representation to
correspond to an operator conjugate to the unperturbed Liouvillian.
We shall carry out our study of  decoherence here in the framework of
the ordinary (direct integral) Hilbert space; the application of these
ideas to the Liouville space will be considered elsewhere.

Any time-dependent observable $A(t)$ defined in the usual quantum
Hilber space can be naturally lifted to the direct integral space
$\bar {\cal H}$ as follows

$$ (\hat A \psi)_t = A(t) \psi_t
\eqno(8)$$
As a natural generalization, the expectation value of any observable
$\hat A$ in the direct integral space is defined by

$$ \langle \hat A \rangle_\psi = {{(\psi, \hat A \psi)_{\bar {\cal H}}}
\over {(\psi, \psi)_{\bar {\cal H}}}} = {{\int dt (\psi_t, A(t)
\psi_t)_{\bar {\cal H}}} \over {\int dt (\psi_t, \psi_t)_{\bar {\cal
H}}}}.
\eqno(9)$$
In what follows, we will formulate and study decoherence in this
framework.

\bigskip
\bigskip\noindent
{\bf 2. Pure and Mixed States}
\bigskip
We wish to show now that a pure state in the larger Hilbert space
(which we will refer to as a pure Lax-Phillips  state), can represent both pure
and mixed states in the usual sense.

Practically, almost all measurement processes correspond to
measurements of observables which are time-dependent in the
Schr\"odinger picture, such as a projection operator to a given
subspace or the asymptotic Heisenberg variables of a  scattering
process.  The lift of such operators to $\bar {\cal H}$ is
$t$-independent also.  Therefore, if two different states give the
same expectation value for a complete set of time independent
observables, the two states are indistinguishable.  In this sense, we
define the following:
\item{1.}A Lax-Phillips  state $\psi$ is called ``pure-like'' if there exists a
pure state $\phi_0$ in the original Hilbert space ${\cal H}$ such that
$$\langle \hat A \rangle_\psi = {(\phi_0, A\phi_0) \over (\phi_0,
\phi_0),}
\eqno(10)$$
holds for any time-independent observable $A$ on the original Hilbert
space ${\cal H}$.

\item{2.} A Lax-Phillips  state $\psi$ is called ``mixed-like'' if there exists
a density matrix $\rho_0$ in the original Hilbert space ${\cal H}$ for which
$Tr \rho^2_0 < 1$, such that
$$\langle \hat A \rangle_\psi = Tr(A \rho_0),
\eqno(11)$$
holds for any time-independent observable $A$ on the original Hilbert
space ${\cal H}$.

For a time-independent obervable $A$ on ${\cal H}$, Eq. (9) gives

$$ \langle \hat A \rangle_\psi = {\int dt (\psi_t,A\psi_t)_{\cal H}
\over \int dt(\psi_t, \psi_t)_{\cal H}} = Tr(\rho_\psi A),
\eqno(12)$$
where $\rho_\psi$ is the density matrix on the original Hilbert space
${\cal H}$ associated with the  Lax-Phillips  state $\psi$, defined as

$$\rho_\psi = {1 \over {\cal N}} \int dt \vert \psi_t \rangle \langle
\psi_t \vert;\qquad {\cal N} = \int dt(\psi_t, \psi_t)_{\cal H}.
\eqno(13)$$
Therefore, if $\rho_\psi$ is pure, the Lax-Phillips  state $\psi$ is pure-like,
and if $\rho_\psi$ is mixed the Lax-Phillips  state $\psi$ is mixed-like.

The state $\psi = \{\psi_t\}$ is pure-like if and only if there is a
normalized state $\psi_0 \in {\cal H}$ and a scalar function $f(t)$
satisfying

$$ \int dt \vert f(t) \vert^2 = 1,$$
such that

$$ \psi_t = \psi_0 f(t).
\eqno(14)$$
Indeed (14) gives ${\cal N}= 1$ and

$$\rho_\psi = \int dt \vert \psi_t \rangle \langle \psi_t \vert = \int
dt \vert f(t) \vert^2 \vert \psi_0 \rangle \langle \psi_0 \vert = \vert
\psi_0 \rangle \langle \psi_0 \vert.
\eqno(15)$$

In general

$$\eqalign{Tr \rho^2_\psi &= \Sigma^i  \int \int dt dt^\prime \langle \psi_i
\vert \psi_t \rangle \langle \psi_t \vert \psi_{t^\prime} \rangle \langle
\psi_{t^\prime} \vert \psi_i \rangle \cr
&= \int \int dt dt^\prime \vert \langle \psi_t \vert \psi_{t^\prime} \rangle
\vert^2,\cr}
\eqno(16)
$$
where $\{\psi_i\}$ is a complete orthonormal set in ${\cal H}$.  By the
Schwartz inequality, unless $\psi_t$ is proportional to $\psi_{t^\prime}$,
i.e., of the form (14),

$$
\vert \langle \psi_t \vert \psi_{t^\prime} \rangle \vert^2 < \Vert \psi_t
\Vert^2_{\cal H} \Vert \psi_{t^\prime} \Vert^2_{\cal H},
\eqno(17)
$$
and hence $Tr \rho^2_\psi < 1$.

\bigskip\noindent
{\bf 3. Decoherence in the Flesia-Piron Approach}
\bigskip

In this and the next sections, we discuss the possibility of
decoherence, or the evolution from pure-like to mixed-like states.
Here we treat the problem in the framework of the Flesia-Piron
approach.

First we consider the Schr\"odinger evolution for a time-dependent
Hamiltonian.  The solution of the time-dependent Schr\"odinger
equation can always be written formally as $\psi_t = U(t, t^\prime)
\psi_{t^\prime}$, where $U(t, t^\prime)$ satisfies the chain property
$U(t, t^\prime)U(t^\prime,t^{\prime\prime}) = U(t,t^{\prime\prime})$,
and can be expressed in
terms of the integral of a time-ordered product.  We define $W_t(\tau)
= U(t+\tau,t)$, and lift the evolution to $\bar {\cal H}$ as follows [9]

$$\psi^\tau_{t+\tau} = W_t (\tau) \psi_t,
\eqno(18)$$
where $W_t(\tau)$ is given by ($T$ implies the time-ordered product)

$$W_t(\tau) = T(e^{-i \int^{t+\tau}_t H(t^\prime) dt^\prime}).
\eqno(19)$$
For this kind of time-evolution we obtain

$$\eqalign{
\rho_\psi (\tau) &= {1 \over {\cal N} (\tau)} \int dt \vert (U(\tau)
\psi)_t \rangle \langle (U(\tau)\psi)_t \vert \cr
&= {1 \over {\cal N}} \int dt \vert (U(\tau) \psi)_{t+\tau} \rangle
\langle (U(\tau) \psi)_{t+\tau} \vert \cr
=& {1 \over {\cal N}} \int dt \vert w_t(\tau) \psi_t \rangle \langle W_t
(\tau) \psi_t \vert, \cr}
\eqno(20)$$
where we have used the fact that the normalization constant is
time-independent (which follows from the unitarity of $U(\tau$)).  For
the pure-like state introduced in (14), we then have

$$ \rho_{\psi_P} (\tau) = \int dt \vert f(t) \vert^2 W_t (\tau) \vert
\psi_0 \rangle \langle \psi_0 \vert W_t^{\dag} (\tau).
\eqno(21)$$
It follows from (16) that this state is ``mixed-like'' if $W_t (\tau)
\vert \psi_0 \rangle$ is not proportional to $W_{t^\prime} (\tau) \vert
\psi_0 \rangle$ for $t \not= t^\prime$ (the set for which $t=t^\prime$
is of zero measure).

The evolution operator $W_t(\tau)$ does not depend on $t$ if and only
if the system is invariant to translations along the $t$-axis, i.e.,
the Hamiltionian $H(t)$ is time-independent.  In this case $W_t(\tau)
= W(\tau) = e^{-iH\tau}$ and

$$ \rho_{\psi P} (\tau) = \int dt \vert f(t) \vert^2 W(\tau) \vert \psi_0
\rangle \langle \psi_0 \vert W^{\dag} (\tau) = W(\tau) \vert \psi_0
\rangle \langle \psi_0 \vert W^{\dag} (\tau)
\eqno(22)$$
is again a pure-state.  In other words, in the Flesia-Piron approach
if the Hamiltonian does not depend on time explicitly, a pure-like
state remains pure-like, and there arises no decoherence.  On the
other hand, if the Hamilitionian depends on time explicitly, the
states, in general, cannot maintain their purity and decoherence takes
place.  As we shall see in a concrete example, the degree of
decoherence depends not only on the time-dependence of the
Hamiltonian, but also on the initial states.
\bigskip
\bigskip\noindent
{\bf 4. Decoherence in Closed System}
\bigskip
As shown in the previous section, the original Flesia-Piron approach
may allow decoherence only for systems with explicit time dependence,
i.e., for open systems.  This is not so satisfactory as the system
plus apparatus can always be seen as a closed system, where
decoherence takes place without external disturbances.  As briefly
explained in the Introduction, the Flesia-Piron approach has been
recently generalized by introducing an interaction which is non-local
on the time axis (cf. eqs. (6) and (7)) [8].  As we shall see now, the
generalization provides a possibility of decoherence even for closed
systems.

Choosing, in particular, a kernel $\kappa$ of the form (cf. eq. (7))

$$\kappa_{t,t^\prime} = \kappa_{t-t^\prime},
\eqno(23)$$
it follows that (here, $-i\partial_t$ stands for the operator on $\bar
{\cal H}$ which is represented as a derivative in the
$t$-representation)

$$ [\kappa, -i\partial_t] = 0.
\eqno(24)$$
Therefore, the system described by generator of this form is closed in
the sense that it is invariant to translations on the time-axis, i.e.,

$$ [K, -i\partial_t] = 0.
\eqno(25)$$
It is shown in the Appendix that this kind of interaction leads to an
evolution operator of the form

$$W_{t,t^\prime}(\tau) = {1 \over 2\pi} \int e^{i(t-t^\prime) \sigma}
e^{-i \kappa(\sigma)\tau} d \sigma = W_{t-t^\prime}(\tau)
\eqno(26)$$
where $\kappa(\sigma)$ is the Fourier transform of
$\kappa_{t-t^\prime}$ with respect to $t-t^\prime$ and is Hermitian.

Now, consider the most general form of pure state, $\psi_t =
f(t)\psi_0$.  If follows from (6) that the time evolution of such a
state is

$$ (\psi^\tau)_{t+\tau} = \int W_{t,t^\prime} (\tau) \psi_0
f(t^\prime) dt^\prime.
\eqno(27)$$
Since the evolution operators satisfy (from (26))
$W_{t,t^\prime}(\tau) = W_{t-t^\prime}(\tau)$, it follows that

$$\eqalign{
(\psi^\tau)_{t+\tau} &= \int W_{t-t^\prime}(\tau) \psi_0 f(t^\prime)
dt^\prime = \cr
&= \int W_{t^\prime}(\tau)\psi_0 f(t-t^\prime) dt^\prime.\cr}
\eqno(28)$$
This corresponds, for every $t$, to a superposition of the states
$W_{t^\prime} (\tau) \psi_0$, but, in general, for each $t$, the
weights are different, and we conclude that the state may be mixed by
the evolution (cf. the arguments below (16)).  The purity of the
state will be conserved if and only if all the states $W_{t^\prime}
(\tau) \psi_0$ are the same up to a factor which is a function of
$t^\prime$ (and $\tau$; the discussion which follows is, however, for
each $\tau$).  We shall now prove that this occurs for any $\psi_0$ if
and only if $\kappa_{t-t^\prime} = \kappa \delta (t-t^\prime)$, where
$\kappa$ is some constant operator.

Let us assume that

$$W_t (\tau) \psi_0 = \alpha_t \psi_1
\eqno(29)$$
for any arbitrary $\psi_0$ and corresponding $\psi_1$.  Let
$\{\phi_n\}$ be a complete orthonormal set in $\bar {\cal H}$; then for
each $\tau$,

$$ W_t(\tau) \phi_n = \alpha_t \psi_n = \alpha_t \sum_n \beta_{mn}
\phi_m,
\eqno(30)$$
and therefore

$$(\phi_m, W_t(\tau) \phi_n) = \beta_{mn} \alpha_t.
\eqno(31)$$
Hence,

$$ W_t(\tau) = \alpha_t W(\tau),
\eqno(32)$$
where

$$ (\phi_m, W(\tau) \phi_n) = \beta_{mn}.
\eqno(33)$$
Taking the Fourier transform of (32) one obtains

$$ \tilde W_\sigma (\tau) = {\tilde \alpha}(\sigma) W (\tau).
\eqno(34)$$
On the other hand, from (26) it follows that

$$ \tilde W_\sigma(\tau) = e^{-i\kappa(\sigma)\tau}.
\eqno(35)$$
We show first that $W(\tau)$ has an inverse.  It is shown elsewhere
[8] that the evolution operators satisfy the relation

$$ \int W_{t,t^{\prime\prime}}(\tau)
W_{t^\prime,t^{\prime\prime}}(\tau)^{\dag} dt^{\prime\prime} =
\delta(t-t^\prime).
\eqno(36)$$
Using (32) if follows from (36) that

$$ \int \alpha_{t-t^{\prime\prime}}
\alpha^\ast_{t^\prime-t{\prime\prime}} dt^{\prime\prime} W (\tau)
W(\tau)^{\dag} = \delta(t-t^\prime),
\eqno(37)$$
and, by integrating it with respect to $t$,

$$ \lambda W(\tau) W(\tau)^{\dag} = 1
\eqno(38)$$
with

$$ \lambda = \int^{+ \infty}_{- \infty} dt \int^{\infty}_{- \infty}
\alpha_{t-t^{\prime\prime}}
\alpha^\ast_{t^{\prime\prime}-t^\prime}dt^{\prime\prime},$$
i.e.,

$$W^{-1}(\tau) = \lambda W(\tau)^{\dag}.
\eqno(39)$$
Hence, $W^{-1}$ exists.  Then, from (34) and (35) it follows that

$$ \tilde W_\sigma (\tau_1) \tilde W_\sigma (\tau_2)^{-1} =
e^{-i\kappa(\sigma) (\tau_1 - \tau_2)} = W (\tau_1) W (\tau_2)^{-1},
\eqno(40)$$
independently of $\sigma$; hence

$$ \kappa(\sigma) = {\rm const.} \Rightarrow \kappa_{t-t^\prime} =
\kappa\delta(t-t^\prime).
\eqno(41)$$
Thus, we realize that pure states remain pure if and only if condition
(41) is satisfied, which is exactly the case of a time-independent,
pointwise Hamiltonian.

The density matrix corresponding to the Lax-Phillips  state (28) is given by

$$\rho_\psi = \int dt \vert \psi^\tau_{t+\tau} \rangle \langle
\psi^\tau_{t+\tau} \vert = \int d\sigma e^{i\kappa(\sigma)\tau} \vert
\psi_0 \rangle \langle \psi_0 \vert e^{-i\kappa(\sigma)\tau},
\eqno(42)$$
with

$$ P(\sigma) = {1 \over 2\pi} \biggl| \int dt e^{-it\sigma} f(t) \biggr|^2.
\eqno(43)$$
Since $P(\sigma)\geq 0$ and $\int d\sigma P(\sigma) = 1$, the
expression (42) implies that the density matrix $\rho_\psi$ is a
convex combination of pure state evolutions by a ``Hamiltonian
$\kappa(\sigma)$" weighted by a ``probability
$P(\sigma)$".  Therefore, it is, in general, a mixed
state.  Moreover, the evolution (42) is formally the same as that
appears in the MHS theory (cf. eq. (5.17) of [4]).  We therefore see
that a generalized evolution of the form (6) may lead to mixing of
pure states for closed systems and that the generalized formulation
may possess all the aspects of the MHS theory.

\bigskip\noindent
{\bf 5. Examples}
\bigskip
In order to see the details of the decoherence processes, we study two
simple examples corresponding to the time-local evolution (4) and the
time-nonlocal evolution (6) respectively.

First, let us consider a system described by the following Hamiltonian

$$ H(t) = -{\Omega_0 \over 2} \Sigma_z + {\Omega \over 2} [\Sigma_+
e^{i\Omega_0 t} + \Sigma_- e^{-i\Omega_0t}],
\eqno(44)$$
where $\Sigma_i$ are the Pauli matrices:

$$ \Sigma_z = \pmatrix{1&0\cr 0&-1\cr}\quad
 \Sigma_+ = \pmatrix{0&1\cr 0&0\cr}\quad
\Sigma_- = \pmatrix{0&0\cr 1&0\cr}
$$
The Hilbert space for this model is the two dimensional complex space
${\cal H} = {\bf C}^2$.  It is easy to derive the evolution operator
$W_t(\tau)$ corresponding to the Hamiltonian $H(t)$.  One obtains

$$ W_t(\tau) = u(\tau) \{\cos {\Omega \over 2} \tau - i \sin {\Omega
\over 2} \tau  (\Sigma_+ e^{i\Omega_0 t} + \Sigma_- e^{-i\Omega_0 t})\},
\eqno(45)$$
where the operator $u(\tau)$ is given by

$$u(\tau) = \pmatrix{\exp(i {\Omega_0 \over 2} \tau) & 0\cr
0 & \exp(-i {\Omega_0 \over 2}\tau).\cr}
\eqno(46)$$

The direct integral space for this model is given by $L^2(-\infty,
\infty; {\bf C}^2)$.  We wish to study now the time evolution of pure-like
state given by Eq. (14).  From Eqs. (18) and (45) we have

$$\eqalign{
\rho_{\psi_P} &= u(\tau) \int dt \vert f(t)\vert^2 \{\cos ({\Omega
\over 2} \tau) - i \sin ({\Omega \over 2} \tau) (\Sigma_+ e^{i\Omega_0t}
+ \Sigma_- e^{-i\Omega_0 t})\} \vert \psi_0 \rangle \cr
&\times \langle \psi_0 \vert \{ \cos({\Omega \over 2} \tau) + i \sin
({\Omega \over 2} \tau) (\Sigma_+e^{i\Omega_0 t}+ \Sigma_- e^{i-\Omega_0
t})\} u^{\dag}(\tau)\cr
&= u(\tau) \biggl[ \cos^2({\Omega \over 2} \tau) \vert \psi_0 \rangle
\langle \psi_0 \vert + \sin^2({\Omega \over 2} \tau) \Sigma_+ \vert
\psi_0 \rangle \langle \psi_0 \vert \Sigma_-\cr
&+ \sin^2 ({\Omega \over 2} \tau) \Sigma_- \vert \psi_0 \rangle \langle
\psi_0 \vert \Sigma_+\cr
&+ \{ iF(\Omega_0) \cos ({\Omega \over 2} \tau) \sin ({\Omega \over 2}
\tau) (\vert \psi_0 \rangle \langle \psi_0 \vert \Sigma_+ - \Sigma_- \vert
\psi_0 \rangle\langle \psi_0 \vert )\cr
&+ \sin^2({\Omega \over 2} \tau) F(2\Omega_0)\Sigma_+ \vert \psi_0
\rangle \langle \psi_0 \vert \Sigma_+ + h.c.\} \biggr] u^{\dag} (\tau),\cr}
\eqno(47)$$
where $F(\omega)$ is the Fourier transform of $\vert f(t)\vert^2$

$$ F(\omega) \equiv \int dt \vert f(t) \vert^2 e^{i\omega t}.
\eqno(48)$$
As an example, suppose we take $\vert f(t) \vert^2$ to be the Gaussian
form

$$ \vert f(t) \vert^2 = {1 \over 2 \sqrt{\pi} \triangle} \exp \biggl[ -
{(t-t_0)^2 \over 4 \triangle^2} \biggr] ,
\eqno(49)$$
then we obtain

$$ F(\omega) = e^{i\omega t_0} e^{-(\omega \triangle)^2}.
\eqno(50)$$

In order to study the decoherence quantitatively, we introduce a
degree of decoherence $\varepsilon$ (which is different from the one
introduced in [3,4])

$$ \varepsilon = Tr \rho^2_{\psi_P} - 1.
\eqno(51)$$
Obviously, $\varepsilon$ represents a ``distance'' from pure states (a
state with  $\varepsilon \not= 0$ is mixed and a state with
$\varepsilon = 0$ is pure).  For the state (47) with the Gaussian form
of $f(t)$ given by (49), a tedious but straightforward calculation gives

$$ \eqalign{
 \varepsilon &= 2 g(\Omega_0) \sin^2 {\Omega\tau \over 2} \{\sin^2
{\Omega\tau \over 2} \vert \langle \Sigma_+\Sigma_-\rangle\vert^2 (\vert
F(\Omega_0 \vert + 1)(\vert F (\Omega_0)\vert^2 +1)(\vert
F(2\Omega_0)\vert + 1)\cr
&+ \cos^2  {\Omega\tau \over 2} ( 1-2 \langle\Sigma_+\rangle\langle
\Sigma_-\rangle )(\vert F(\Omega_0)\vert + 1) + \biggl[ \cos^2  {\Omega\tau
\over 2} \langle \Sigma_+\rangle^2 F(\Omega_0)^2 (\vert F(\Omega_0)\vert
+ 1)\cr
&+ i \sin  {\Omega\tau \over 2}\cos  {\Omega\tau \over 2}
\langle\Sigma_-\rangle \langle\Sigma_z\rangle F(\Omega_0)^\ast (\vert
F(\Omega_0)\vert + 1)(\vert F(\Omega_0)\vert^2 +1) + {\rm c.c.}\biggr] \}, \cr}
\eqno(52)$$
where $\langle\dots\rangle$ stands for the average with respect to the
state $\psi_0$ and the function $g(\Omega_0) \equiv \vert
F(\Omega_0)\vert - 1$ describes the initial state dependence of the
degree of decoherence.  The deviation of the state $\rho_{\psi_P}$
from pure states can also be seen directly on the operator level.
Indeed, it can be rewritten as

$$ \eqalign{
\rho_{\psi_P} &= u(\tau) \tilde W (\tau)
\vert \psi_0 \rangle\langle\psi_0 \vert \tilde W^{\dag} (\tau) u^{\dag}
(\tau)\cr
&+ g(\Omega_0) \sin {\Omega(\tau)\over 2} u(\tau) \{ i \ e^{i\Omega_0
t_o} \cos {\Omega(\tau)\over 2} (\vert \psi_0 \rangle\langle\psi_0
\vert \Sigma_+ - \Sigma_+ \vert \psi_0 \rangle\langle\psi_0 \vert)\cr
&+ \sin {\Omega(\tau)\over 2} (\vert F (\Omega_0) \vert +1)(\vert F
(\Omega_0) \vert^2 +1) e^{2i\Omega_0 t_0}\Sigma_+ \vert \psi_0
\rangle\langle\psi_0 \vert \Sigma_+ + {\rm h.c.}\}u^{{\dag}(\tau)},\cr}
\eqno(53)$$
where $\tilde W (\tau)$ is a unitary operator given by
$$ \tilde W(\tau) = \cos {\Omega(\tau)\over 2} - i\sin
{\Omega(\tau)\over 2} (e^{i\Omega_0 t_0} \Sigma_+ + e^{-i\omega_0 t_0}
\Sigma_-).
\eqno(54)$$
Strictly speaking, as the degree of coherence is different from zero
$(\varepsilon \not= 0)$, decoherence takes place irrespective of the
value of $\triangle(\not= 0)$.  However, if $g$ and thus the degree of
decoherence $\varepsilon$ are very small, the state $\rho_{\psi_P}$
corresponds to an almost pure state.  In short, we find for the
Gaussian example in the Flesia-Piron approach, that when the initial
state is well localized on the t-axis compared with the time scale of
the change of the Hamiltonian, i.e., $\Omega_0 \triangle << 1$, the
state $\rho_{\psi_P}$ remains practically pure.  Otherwise,
decoherence takes place.

Next we consider a system with a time-nonlocal interaction:

$$ \kappa_{t,t^\prime} = \kappa_{t-t^\prime} = {\Omega \over 2}
(\Sigma_+ \delta(t-t^\prime + t_d)+ \Sigma_- \delta(t-t^\prime - t_d),
\eqno(55)$$
which leads to

$$ \kappa(\sigma) = \int d(t-t^\prime) e^{-i(t-t^\prime)\sigma}
\kappa_{t-t^\prime} = {\Omega \over 2} (\Sigma_+ e^{i t_d \sigma} +
\Sigma_- e^{-i t_d \sigma}),
\eqno(56)$$
and, thus,

$$ e^{-i\kappa(\sigma)\tau} = \cos {\Omega\tau\over 2} - i \sin
{\Omega\tau\over 2}  (e^{i\sigma t_d}\Sigma_+ + e^{-i\sigma t_d}
\Sigma_-).
\eqno(57)$$
Because of (42), we then have

$$ \eqalign{
\bar \rho_{\psi_P} &= \int d\sigma P(\sigma) \{\cos {\Omega\tau\over
2} - i \sin {\Omega\tau\over 2} (e^{i\sigma t_d} \Sigma_+ +
e^{-i\sigma t_d}\Sigma_-)\}\vert \psi_0\rangle\cr
&\times \langle \psi_0\vert \{\cos{\Omega\tau\over 2} + i \sin
{\Omega\tau\over 2} (e^{i\sigma t_d} \Sigma_+ + e^{-i\sigma
t_d}\Sigma_-)\}\cr
&= \cos^2{\Omega\tau\over 2}\vert \psi_0 \rangle\langle\psi_0 \vert
+ \sin^2 {\Omega\tau\over 2} \Sigma_+ \vert \psi_0
\rangle\langle\psi_0 \vert \Sigma_-\cr
&+ \sin^2{\Omega\tau\over 2} \Sigma_-\vert \psi_0 \rangle\langle\psi_0
\vert \Sigma_+\cr
&+ \{ i \bar F (t_d) \sin {\Omega\tau\over 2} \cos{\Omega\tau\over
2} (\vert \psi_0 \rangle\langle\psi_0 \vert\Sigma_+ - \Sigma_+ \vert
\psi_0 \rangle\langle\psi_0 \vert) \cr
&+ \sin^2 {\Omega\tau\over 2} \bar F (2t_d) \Sigma_+ \vert \psi_0
\rangle\langle\psi_0 \vert \Sigma_+ + {\rm h.c.}\}, \cr}
\eqno(58)$$
where $\bar F(t)$ is the Fourier transform of the ``probability
density'' $P(\sigma)$,

$$\bar F(t) \equiv \int d\sigma P(\sigma) e^{i\sigma t}.
\eqno(59)$$

{}From eqs. (47) and (58), we find that the density operators
$\rho_{\psi_P}$ and $\bar \rho_{\psi_P}$ have an identical form except
for the unitary operator $u(\tau)$ and the fact that $F$ in (47) is
replaced by $\bar F$ in (58).   Thus, their decoherence properties are
the same.  For the Gaussian form (49) of $F(t)$, we have

$$\bar F(t) \ \exp(-{t^2 \over 16\triangle^2}) = \vert
F(t/(4\triangle^2))\vert.
\eqno(60)$$
Therefore, in this case, the degree of decoherence $\varepsilon$ and
the decomposition into purity preserving and purity nonpreserving
terms of the density operator $\bar \rho_{\psi_P}$ can be obtained
from the corresponding expressions (52) and (53) for $\rho_{\psi_P}$
by replacing $\Omega_0$ with $t_d/(4\triangle^2)$ and dropping
$e^{i\Omega_0 t_0}, e^{2i\Omega_0 t_0}$ and $u(\tau)$. We then find
that when the initial state is well delocalized on the t-axis compared
with the non-locality of the interaction, i.e., $t_d/\triangle << 1$,
the state $\rho_{\psi_P}$ remains practically pure.

Interestingly enough, the dependence of the degree of decoherence
$\varepsilon$ upon the initial spread $\triangle$ on the t-axis for
the first example is opposite to that for the second one.  The
difference can be understood as follows:  In the first example, as the
Hamiltonian changes in time, the state with larger spread cannot
follow the change of the Hamiltonian and loses its coherence.
Contrarily, in the second example, the state can preserve its purity
only when it has enough spread not to feel the non-locality of the
interaction.

In short, we have shown that the details of the decoherence proceses
depend not only on the dynamics of the systems but also on the initial
conditions.

\bigskip\noindent
{\bf 6. Conclusion}
\bigskip
We have that the Lax-Phillips  theory provides a description of the quantum
states which admits the possibility of decoherence for time-dependent
Hamiltonian systems, and even for systems which are closed (but not of
Hamiltonian form in the  original Hilbert space).  As we have seen in
the concrete examples, the degrees of decoherence depends not only on
the structure of the system (i.e., the generator of motion), but also
on the initial conditions.  In other words, a given system may behave
as a (almost) pure quantum system (coherent time evolution) or as a
system plus an apparatus (incoherent time evolution) depending on the
initial conditions.  Intermediate situations are also possible.
Therefore, the Many-Hilbert-Space theory can be formulated naturally
in this framework, and it is not necessary to specify the limit
between the system and the measuring apparatus.  As we have remarked,
the relation between the singularities of the $S$-matrix and the
spectrum of the generator of the semigroup can be obtained only from
such a general evolution.  We therefore see that the origin of
irreversibility may be found in such structures.

\bigskip\noindent
{\bf 6. Acknowlegements}
\smallskip
One of us (LPH) wishes to thank I. Prigogine for his hospitality
during his recent visits to Brussels, and for continuing interest and
encouragement and many helpful and stimulating comments.  He is also
grateful to S.L. Adler for his hospitality at the Institute for
Advanced Study, where his work was partially  supported by the
Monell Foundation.  One of us (ST) is grateful to Prof. I. Prigogine
for his continuous interest and encouragement as well as fruitful
discussions on various problems, to which several ideas of the present
work are owed, and to Prof. M. Namiki for helpful discussions and
information on the Many-Hilbert-Space theory.  We are also grateful to
I. Antoniou for very helpful and fruitful discussions.

It is a pleasure to dedicate this paper to Constantin Piron.  We have
all learned very much from his works on the beauty of the ideas of
physics and their precise mathematical formulation.  One of us (LPH),
in particular, wishes to express his deep appreciation for many years
of warm friendship and for the stimulation and inspiration, both human
and professional, that this relationship has brought to him.

\vfill
\eject
\bigskip\noindent
{\bf Appendix}
\bigskip
The basic equation which relates the generator $\kappa$ to the evolution
operators is as follows:

$$ i\partial_\tau W_{t,t^\prime} (\tau) = \int \kappa_{t+\tau,t^{\prime\prime}}
W_{t^{\prime\prime},t^\prime} (\tau) dt^{\prime\prime}.
\eqno(A1)$$
Let us take the Fourier transform of (A1) with respect to $t$ and $t^\prime$.

$$ i\partial_\tau W_{\sigma,\sigma^\prime} (\tau) = \int
e^{i\sigma\tau} \kappa_{\sigma,\sigma^{\prime\prime}}
e^{-i \sigma^{\prime\prime}\tau} W_{\sigma^{\prime\prime},\sigma^\prime}
(\tau) d\sigma^{\prime\prime}.
\eqno(A2)$$
For closed systems, in which $\kappa_{t,t^\prime} = \kappa_{t-t^\prime}$,
one obtains

$$\kappa_{\sigma,\sigma^\prime} = \tilde \kappa (\sigma) \delta
(\sigma - \sigma^\prime).
\eqno(A3)$$
Using (A3) in (A2), one obtains

$$i\partial_\tau W_{\sigma,\sigma^\prime} (\tau) = \tilde \kappa (\sigma)
W_{\sigma, \sigma^\prime} (\tau)
$$
from which follows

$$ W_{\sigma,\sigma^\prime} (\tau) = e^{-i\tilde\kappa (\sigma)\tau}
\delta(\sigma-\sigma^\prime).
\eqno(A4)$$
Taking the inverse transform of (A4) we get

$$\eqalign{
W_{t,t^\prime} (\tau) &= {1 \over 2\pi} \int d\sigma d\sigma^\prime
e^{i\sigma t} e^{-i \tilde\kappa (\sigma) \tau} \delta(\sigma-\sigma^\prime)
e^{-i \sigma^\prime t^\prime} =\cr
&= {1 \over 2\pi} \int d\sigma e^{i\sigma(t-t^\prime)} e^{-i \tilde\kappa
(\sigma)\tau} \cr}
\eqno(A5)$$
which depends only in $t-t^\prime$.
\vfill
\eject
\bigskip\noindent
{\bf References}
\smallskip
\frenchspacing
\item{1.} For a review, see {\it Quantum Theory and Measurement},
ed. J.A. Wheeler and W.H. Zurek, Princeton University Press, Princeton,
1993.
\item{2.} S. Machida and M. Namiki, {\it Prog. Theor. Phys.} {\bf 63},
1457 (1980); {\bf 63}, 1833 (1980) see also, M. Namiki, {\it Found. Phys.}
{\bf 18} 29 (1988).
\item{3.} M. Namiki and S. Pascazio, {\it Phys. Lett.} {\bf A147} 430 (1990);
{\it Phys. Rev} {\bf A44} 39 (1991).
\item{4.}  M. Namiki and S. Pascazio, ``Quantum Theory of Measurement,"
{\it Phys. Rep.}, in press, (1993) and references therein.
\item{5.} P.D. Lax and R.S. Phillips, {\it Scattering Theory}, Academic
Press, New York, 1967.
\item{6.} C. Flesia and C. Piron, {\it Helv. Phys. Acta} {\bf 57}
697 ( 1984).
\item{7.} L.P. Horwitz and C. Piron, {\it Helv. Phys. Acta}, in press.
\item{8.} E. Eisenberg and L.P. Horwitz, {\it Adv. Chem.
Phys.}, to be published.
\item{9.} We thank I.M. Sigal for a discussion of this procedure in the
context of Floquet theory.
\item{10.} C. George, {\it Physica} {\bf 65} 277 (1973); I. Prigogine,
C. George, F. Henin and L. Rosenfeld, {\it Chemica Scripta} {\bf 4} 5
(1973); T. Petrosky and I. Prigigone, {\it Physica} {\bf A175} 146 (1991);
I. Prigigone, Proc. of ``Quantum Physics and the Universe," eds. M. Namiki
et al. ({\it Vistas in Astronomy} {\bf 37} 7 (1993)) and references
therein.

\end
\bye